\documentclass[aps,prl,twocolumn,showpacs,groupedaddress]{revtex4}
\usepackage{graphicx}
\usepackage{dcolumn}
\usepackage{amsmath}
\begin{document}
\title{Evidence for a New Dissipationless Regime in 2D Electronic Transport}
\author{M. A. Zudov}
\author{R. R. Du}
\affiliation{Department of Physics, University of Utah, Salt Lake City, Utah 84112}
\author{L. N. Pfeiffer}
\author{K. W. West}
\affiliation{Bell Laboratories, Lucent Technologies, Murray Hill, New Jersey 07974}
\received{1 October 2002}
\begin{abstract}
In an ultra-clean 2D electron system (2DES) subjected to crossed millimeterwave (30--150 GHz) and weak ($B < 2$ kG) magnetic fields, a series of apparently dissipationless states emerges as the system is detuned from cyclotron resonances.
Such states are characterized by an exponentially vanishing low-temperature longitudinal resistance and a classical Hall resistance.
The activation energies associated with such states exceeds the Landau level spacing by an order of magnitude.
Our findings are likely indicative of a collective ground state previously unknown for 2DES.
\end{abstract}
\maketitle

Scattering and dissipation are central issues in quantum transport in electronic systems.
The integer and the fractional quantum Hall effects (QHE) \cite{qhe97} are the most remarkable dissipationless transport phenomena observed in a two-dimensional electron systems (2DES), when subjected to strong perpendicular magnetic fields ($B$) and low temperatures ($T$).
Such effects arise from the opening of energy gaps, with the underlying states related to the Landau level (LL) filling factor, $\nu= hn_e/eB$ ($n_e$ is an electron density).
While the integer QHE can be understood in terms of the electron localization in LL gaps, at integer $\nu$, the fractional QHE stems from the formation of an incompressible quantum fluid with fractionally charged quasiparticles, at rational $\nu$.
In standard transport experiments, QHE exhibits an exponentially vanishing longitudinal resistance, $R_{xx}\rightarrow 0$, and a quantized Hall resistance, $R_{xy}=(1/\nu)h/e^2$, as $T\rightarrow 0$.

The vanishing of $R_{xx}$ in the QHE regime indicates complete suppression of scattering of electrons, much the same as in a case of charge superfluidity.
Novel dissipationless effects have been proposed to exist, theoretically, in 2D electronic transport regimes other than the QHE.
For example, Fr\"ohlich superfluidity, the transport of a phase-coherent sliding charge-density wave \cite{cdw}, is an ideal dissipationless effect.
Superfluidity of excitonic systems, such as those realized by optical pumping \cite{keldysh} or in a coupled two-layer electron-hole quantum well (QW) \cite{lozovik, fukuzawa} have also been long-sought-after but remained experimentally elusive.
More recently, excitonic superfluidity has been reported for a double electronic QW \cite{spielman}.
Observation of dissipationless transport in a new experimental regime, in addition to QHE, is of fundamental importance to correlation physics since it signals a novel ground state.

Until recently, magnetotransport in 2DES under the influence of a millimeterwave (MW) field has often revealed photoconductivity related to electron spin resonance \cite{dobers88} or long-wavelength magnetoplasmons \cite{vas93}, with a signal level of a few percent.
With increased 2DES mobility, giant amplitude oscillatory magnetoresistance was discovered by Zudov et al.\ \cite{zudovprb,zudov1} using MW excitation.
While the underlying mechanism for such effect has not been completely understood, the period (in $1/B$) of the oscillations is believed to be controlled by the ratio of the MW frequency, $\omega$, to the cyclotron frequency, $\omega_c=eB/m^*$ ($m^*=0.068\,m_0$ is the effective mass of band electrons in GaAs).
Specifically, the $B$-positions of maxima and minima in the oscillatory structure were found to conform to:
\begin{equation}
\frac{\omega}{\omega_c} \equiv \varepsilon = \left\{
\begin{array}{ll}
j, & {\rm maxima} \\
j+1/2, & {\rm minima}
\end{array}
\right., \,\,\,\, j=1,\,2,\, 3,\,...\,,
\label{crexp}
\end{equation}
where $j$ is the difference between the indices of the participating LLs.
Phenomenologically, such oscillations resemble Shubnikov-de Haas (SdH) oscillations except that their period relates to $\varepsilon$ rather than $\nu$.
Similar oscillatory $R_{xx}$, but with an even greater amplitude, has been observed in transmission-line experiments \cite{engel}.
Most remarkably, both experiments have reported that at the minima ($\varepsilon = j+1/2$), the $R_{xx}$ was {\em reduced} as compared to its dark value indicating a suppression of scattering events for electron transport.
Increasing sample mobility favors oscillation amplitude, and, therefore, the minima become progressively stronger under similar experimental conditions.
Intuitively, one would wonder if the minima ultimately approach zero in a very clean 2DES.

In this paper we report on experimental evidence for dissipationless transport related to these minima observed in an ultra-clean GaAs/Al$_x$Ga$_{1-x}$As QW sample.
As the system is detuned from cyclotron resonance conditions ($\varepsilon=j$), a series of deep $R_{xx}$ minima emerges at low temperatures.
Transport in such states is characterized by an exponentially vanishing $R_{xx}$ and a largely classical Hall resistance, as $T \rightarrow 0$.
For stronger minima, the $R_{xx}$ values are indistinguishable from zero within experimental uncertainty.
Temperature-dependent $R_{xx}$ shows an activated behavior with an energy scale exceeding the LL spacing by an order of magnitude.

While the origin of such extraordinary transport behavior is presently unclear, the electron-electron correlations appear to be responsible for such a complete suppression of scattering.
Our observation is likely pointing toward a novel, many-electron condensed phase induced by MW excitation.

\begin{figure}[tbp]
\resizebox{0.47\textwidth}{!}{
\includegraphics{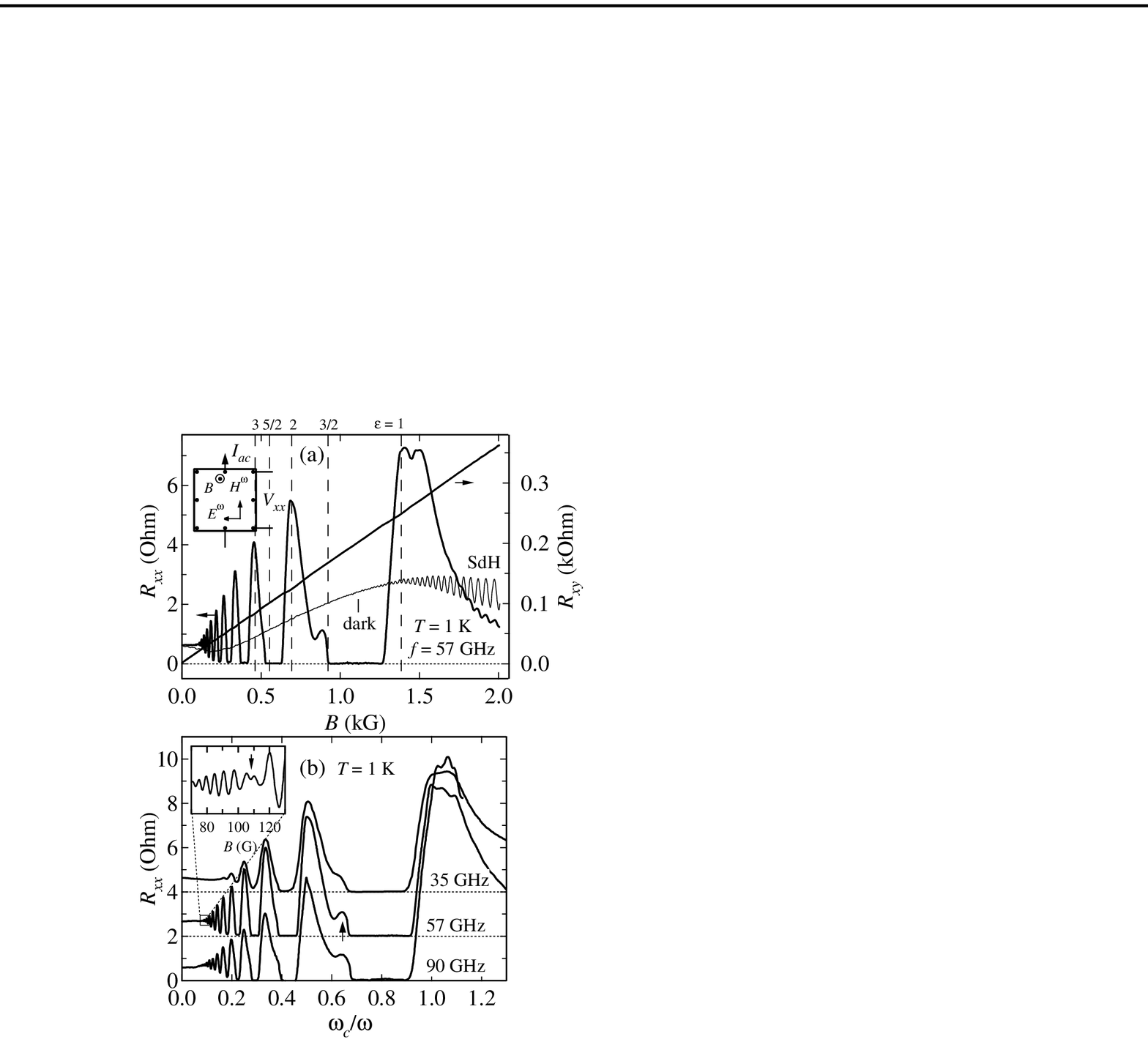}
}
\caption{
(a) Thick lines represent $R_{xx}$ (left axis) and $R_{xy}$ (right axis) under MW ($f=57$ GHz) illumination at $T\approx 1$ K.
Thin line shows dark $R_{xx}$.
Vertical dashed lines were calculated using Eq.\ (\ref{crexp}) and are marked by $\varepsilon$ on top axis.
Inset depicts sample layout, typical electrical connections, and orientations of current $I$, magnetic field $B$, electromagnetic wave fields $E^\omega$ and $H^\omega$.
(b) $R_{xx}$ ($T\approx 1$ K) under MW illumination for $f$ = 35, 57, and 90 GHz plotted against the $\omega_c/\omega$ (vertically offset for clarity).
The vertical arrow shows the secondary peak.
Inset amplifies the low-$B$ part of the $f$ = 57 GHz trace showing the beats at $B \lesssim 0.1$ kG.
}
\label{fig1}
\end{figure}

Our samples \cite{samples} were cleaved from a Al$_{0.24}$Ga$_{0.76}$As/ GaAs/Al$_{0.24}$Ga$_{0.76}$As QW wafer grown by molecular beam epitaxy.
The width of the QW was 300 \AA $\,$ and the electrons were provided by two Si sheets placed symmetrically above and below the QW at 800 \AA.
In this experiment the mobility, $\mu$, and the electron density, $n_e$ were $2.5 \times 10^7$ cm$^2$/Vs and $3.5 \times 10^{11}$ cm$^{-2}$, respectively.
Such parameters were obtained by a brief illumination from a red light-emitting diode at $T\approx 1.5$ K.
The specimens were $\sim 5$ mm $\times$ 5 mm squares with eight In contacts diffused along the perimeter.
The experiment was performed in a $^3$He cryostat equipped with a superconducting magnet; the MW setup was of Faraday geometry (cf.\ inset, Fig.\ \ref{fig1}(a)).
Coherent, linearly polarized MWs were provided by a set of Gunn diodes (typical power is 10 mW to 20 mW, varied by an attenuator) covering 30--150 GHz (1.5--7.5 K) frequency range.
The radiation was sent down to the experiment via an oversized (WR-28) waveguide.
The specimen was immersed in the $^3$He coolant kept at a constant temperature controllable from 0.5 K to 6 K and monitored using a calibrated carbon glass thermometer.
$R_{xx}$ and $R_{xy}$ were measured employing low-frequency lock-in technique, in sweeping $B$, at constant $T$, and under {\em continuous} MW illumination of fixed frequency and power \cite{power}.

In Fig.\ \ref{fig1}(a) we present typical data for the $R_{xx}$ (left axis) and the $R_{xy}$ (right axis) taken at $T\approx 1$ K under $f\equiv \omega/2\pi=57$ GHz illumination.
For comparison, we also plot the dark $R_{xx}$ (thin line) which reveals only regular SdH oscillations at $B \gtrsim 1.2$ kG.
While both the $R_{xx}$ curves converge as $B \rightarrow 0$, they are drastically different at $B \gtrsim 50$ G, as the sharp oscillatory structure emerges under MW illumination.
The peaks are dramatically stronger than those in the $\mu=3.0 \times 10^6$ cm$^2$/Vs samples previously reported \cite{zudov1,zudovprb}; the resonant ($\varepsilon = j$) $R_{xx}$ is enhanced up to five times due to MWs.
However, the most striking feature of the $R_{xx}$ data is the emergence of wide, apparently {\em zero-resistance} (within experimental uncertainty) regions punctuated only at the intervals of those peaks.
In what follows we shall examine the $R_{xx}$ and $R_{xy}$ structures in the 0.3--2.0 kG $B$-range, where such features are the strongest.

Vertical dashed lines in Fig.\ \ref{fig1}(a) were drawn in accordance with Eq.\ (\ref{crexp}) for $j$ = 1, 2, 3, and $f=57$ GHz.
While the peak positions can be rather accurately described by Eq.\ (\ref{crexp}) for $\varepsilon=j$, their shapes are largely asymmetric, with a steeper left (lower-$B$) slope.
The zero-resistance regions in such ultra-clean samples are {\em not centered} around the $\varepsilon=j+1/2$; they reside primarily on the higher-$B$ side from those half-integers.
In fact, such half-integers (3/2, 5/2, and 7/2) seem to represent a sharp boundary between the $R_{xx}$ peaks and the neighboring minima.
We interpret the dramatic shift of the minima positions with respect to Eq.\ (\ref{crexp}) as an indication of modified energy spectrum due to electron-electron interactions, prevailing in ultra-clean 2DES.

The widths of the zero-resistance regions, furthermore, encompass a large range of the filling factor $\nu$.
For instance, the strongest of such regions, 0.94 kG $<B<$ 1.27 kG, corresponds to a range of $155\gtrsim \nu \gtrsim 115$ in this sample.
Roughly the same widths ($\Delta \nu \approx 40$) were obtained for each of the other major minima.
This fact reinforces our earlier conclusion \cite{zudovprb} that the electronic transport here is controlled by $\varepsilon$ rather than by $\nu$.
Notice that the SdH is absent in this weak $B$ regime, indicating that LLs are strongly mixed and cannot be resolved at a macroscopic scale.
The electron scattering time {\em between} LLs (especially for spin-related events), on the other hand, are found to be exceedingly long \cite{dobers88, zudovprb}, thereby resulting in sharp $R_{xx}$ structures.

Despite the drastic behavior in the $R_{xx}$, little change has been observed in the Hall resistance, $R_{xy}$.
In $B<$ 2 kG, we have observed a classical $R_{xy}$ trace, i.e., $R_{xy}=eB/n_e$, with $n_e \cong 3.5 \times 10^{11}$ cm$^{-2}$.
This value is consistent within 1\% with the $n_e$ obtained from the SdH and the QHE at higher $B$, indicating that electrons are largely delocalized in this regime.
While deviations from the linearity under MWs are noticeable, they appear to accompany only the $R_{xx}$-maxima but not the zero-resistance regions, and thus cannot be viewed as the precursors of the Hall plateaus.
In fact, these deviations can be explained quantitatively by a trivial coupling between the components of the resistivity tensor, since an increment in the $R_{xx}$, $\delta R$, translates into a correction of $-\delta R/(\omega_c\tau)$ in the $R_{xy}$ \cite{coleridge89}.
The absence of Hall plateaus rules out a scenario of the MW-assisted carrier localization and, combined with the apparent independence on $\nu$, presents the main observational distinction from the QHE.
Our findings thus represent, unequivocally, a new dissipationless transport regime in a clean 2DES.

In Fig.\ \ref{fig1}(b) we show additional details of our observations.
First, we have repeated the experiment using different $f$ and obtained similar results.
Typical traces for three selected frequencies of 35, 57, and 90 GHz are plotted against the normalized magnetic field axis, i.e., $1/\varepsilon = \omega_c/\omega$.
Plotted in such a way, these traces largely collapse together except for a progressively better resolution at higher $f$; such data demonstrates that the observed transport behavior is generic over a wide range of frequencies.
Second, at all frequencies we have detected secondary peaks appearing on the higher-$B$ shoulders of the main peaks.
This is best observed for $j=2$ (cf.\ vertical arrow in Fig.\ \ref{fig1}(b)) but is also present for higher orders.
Note that such peaks are consistently observed in large square samples and therefore cannot be explained by magnetoplasmon resonances previously seen in narrow Hall-bar samples \cite{vas93, zudovprb}.
The $B$ positions of such peaks shift concurrently with the main peaks, and thus do not exhibit the characteristic dispersion of the plasmon modes.
Third, a close inspection of the data in lower-$B$ range (cf.\ inset in Fig.\ \ref{fig1}(b)) reveals beats in $R_{xx}$; for example, for $f=57$ GHz the oscillations disappear at $\varepsilon \approx 12$ but then reappear with the opposite phase and persist up to $\varepsilon \approx 22$ (down to $B_0 \approx 65$ G).
Such beats were also observed at other frequencies.


\begin{figure}[tbp]
\resizebox{0.47\textwidth}{!}{
\includegraphics{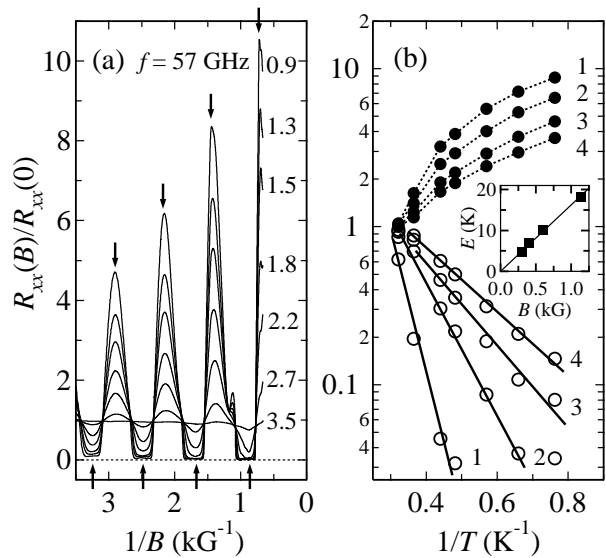}
}
\caption{
(a) $R_{xx}(B)/R_{xx}(0)$ under MW ($f=57$ GHz) illumination, plotted vs. $1/B$ at different $T$ from 0.9 K to 3.5 K.
Upward arrows mark positions of the minima suggesting a roughly periodic series in $1/B$.
(b) $R_{xx}(B)/R_{xx}(0)$ at the minima series (open circles) and at the $j$ = 1, 2, 3, and 4 maxima series (solid circles), as a function of $1/T$.
Solid lines are fits to $R_{xx}(T) \propto $ exp$(-E/T)$.
The inset shows the values of the fitting parameter $E$ as a function of $B$ revealing roughly linear dependence with a slope of $\simeq 18$ K/kG.
}
\label{fig2}
\end{figure}

As much as the $R_{xx}$ is concerned, transport at the minima is remarkably reminiscent of the QHE.
In Fig.\ \ref{fig2}(a) we plot $R_{xx}(B)/R_{xx}(0)$ traces versus $1/B$ (rather than $B$), recorded at a constant $T$ (from 0.9 K to 3.5 K), constant MW power, and $f=57$ GHz.
As typically observed in the QHE regime, zero-resistance regions first become narrower and eventually disappear with increasing $T$; the peaks diminish as well.
Note that while the oscillatory structure disappears in the high-$T$ limit ($T \gtrsim 4$ K), the $R_{xx}(B)$ is not restored to its dark value; it becomes essentially flat (more so below $B\approx$ 2 kG) as opposed to the dark trace in Fig.\ \ref{fig1}(a).

At the intermediate $T$, the minima become sharply-defined, allowing us to measure their $B$-positions with a refined resolution.
Surprisingly, the first few minima seem to form their own sequence, roughly periodic in $1/B$, as marked by the upward arrows in Fig.\ \ref{fig2}(a).
The period of the minima sequence is about 20\% larger than that of the peaks.
We may, without resorting to a specific theoretical model, take this as a spectroscopy evidence that such minima occur at energy intervals about 20\% blue-shifted from the cyclotron transitions $j\hbar\omega_c$.

Using standard Arrhenius plot, we present in Fig.\ \ref{fig2}(b) the $R_{xx}/R_{xx}(0)$ of the first four minima (open circles) and maxima (filled circles) on a logarithmic scale, versus $1/T$.
As a rough estimate, the uncertainty for the $R_{xx}$ measured is $\pm 0.01$ $\Omega$.
All four minima conform to a general expression, $R_{xx}(T)\propto$ exp$(-E/T)$, over at least one decade in $R_{xx}$.
We therefore conclude that the resistance in these regions vanishes exponentially with decreasing $T$.

Quantitatively, an unusually large activation energy appears to associate with such minima.
For example, for the first minimum we extract $E \simeq 20$ K.
This value exceeds both the MW energy ($\hbar \omega \simeq 3$ K) and the LL spacing ($\hbar \omega_c \simeq 2$ K) by nearly an order of magnitude.
Furthermore, by deducing the $E$ at other minima, we observe a rough linearity between $E$ and $B$ with a slope of $\simeq$ 18 K/kG (cf.\ inset in Fig.\ \ref{fig1}(b)).
Overall, the $T$-dependence of the $R_{xx}$ in this regime cannot be reconciled with any simple model based on a single-electron spectrum in a weak magnetic field.

One related question arises concerning the fate of the $R_{xx}$ minima at the lowest $T$.
The exponential $T$-dependence may be taken, as in the case of QHE, as the critical evidence for a complete dissipationless transport extrapolated to $T=0$.

We cannot, however, rule out the possibility for a phase transition, characterized by $T_c$, below which the system becomes dissipationless even at finite $T$.
The prospect for observation of the $T_c$, possibly related to Kosterlitz-Thouless transition \cite{kost} in this system is extremely interesting, but ultimately, such issues must be addressed by further experiments.

In conclusion, we have reported for the first time the evidence for a new dissipationless transport regime, experimentally observed in an ultra-clean 2DES in crossed MW and weak $B$ fields.
By varying a single experimental parameter, namely the mobility of the 2DES, we find a dramatic transition from SdH-like MW photoconductivity oscillations in a moderate-mobility 2DES to zero-resistance states in an ultra-clean 2DES.

The nature of such extraordinary electronic transport is presently unclear, but may be related to the magnetoexcitations and their condensation in the 2DES created by MW excitations.
There exist in this system at least three types of fundamental processes that need to be taken into consideration.
The first is the multiple cyclotron transitions between LLs separated by $j\hbar\omega_c$, which can be identified from the $B$-positions of the strong $R_{xx}$ peaks observed previously \cite{zudov1,zudovprb}, and confirmed again in this experiment.
The second is spin-dependent cyclotron transitions, with energy shifts due to Zeeman splitting in a weak magnetic field.
Such transitions become possible due to spin-orbital coupling effects, but are difficult to identify because the bare $g$-factor in GaAs is rather small.
However, it remains to be seen whether or not the beats seen in a very weak $B$ are in effect manifesting such transitions.
And third, the long-wavelength plasmon modes are believed to disperse into the cyclotron resonance, because their energies are much too low in this sample.
2DES in a weak magnetic field, where many LLs are occupied, is known to exhibit unique properties due to modified electron-electron correlations \cite{aleiner}.
In particular, a 2D electron liquid must be favored in the ultra-clean systems.
Low-energy (on the MW scale) magnetoexciton formed between the excited electron and the hole created in the liquid, in a weak magnetic field, is an attractive candidate for considering the mechanism related to dissipationless transport.
In principle, 2D excitonic condensates could exhibit superfluidity at low temperatures.
Possible excitations of spin-degree of freedom, such as Skyrmion-like entities, would be interesting to examine in the content of our experiments.

We thank C. L. Yang and J. Zhang for assistance in sample preparation and experiments.
Valuable discussions with D. C. Tsui, H. L. Stormer, J. A. Simmons, L. W. Engel, J. M. Worlock, A. L. Efros, Y. S. Wu, and A. A. Koulakov are gratefully acknowledged.
This work (M.A.Z. and R.R.D.) is supported by NSF and by DARPA. R.R.D. is also supported by A. P. Sloan Foundation.




\begin{thebibliography}{10}
\bibitem{samples}
 \bibinfo{note}{Similar results were also obtained in samples from
 GaAs/ Al$_{0.25}$Ga$_{0.75}$As heterostructure wafers, and from both the square and Hall bar geometries.}
\bibitem{power}
 \bibinfo{note}{The MW power density on 2DES is not accessible in this
 experiment; it is reasonable to believe that it is comparable with the values
 measured in transmission-line experiments \cite{engel}. This issue will be
 addressed in a separate paper.}
\end{thebibliography}

\end{document}